\documentclass[
    reprint,
    amsmath,
    amssymb,
    aps,
    prl,
    letter,
    longbibliography,
]{revtex4-1}

\usepackage{graphicx}
\usepackage{booktabs}
\usepackage{amsfonts}
\usepackage{amsmath}
\usepackage{amssymb}
\usepackage{natbib}
\usepackage{url}
\usepackage{hyperref}
\usepackage{subcaption}
\hypersetup{colorlinks=false,pdfborder={0 0 0}}
\usepackage{float}

\usepackage[utf8]{inputenc}
\usepackage{color}
\usepackage[T1]{fontenc}
\usepackage[english]{babel}

\newcommand{\Sil}{$\textrm{SiO}_2$}
\newcommand{\Gromacs}{\textsc{Gromacs~2016}}

\newcommand{\etal}{\textit{et~al.~}}

\newcommand{\qO}{q_\textrm{O}}
\newcommand{\qSi}{q_\textrm{Si}}
\newcommand{\clDiss}{\dot{E}_{\mu_f}}
\newcommand{\eneGainRate}{\dot{E}_\gamma}
\newcommand{\clForce}{F_\textrm{cl}}
\newcommand{\fricCoeff}{\hat{\mu}_f}
\newcommand{\eneBarrier}{\Delta E}
\newcommand{\thetadiff}{\left ( \mydeg{90} - \theta \right )}

\newcommand{\pow}[2]{#1 \, \cdot \, 10^{#2}}
\newcommand{\mydeg}[1]{#1^\circ}

\begin{document}

\begin{abstract}
A hydrophilic liquid, such as water, forms hydrogen bonds with a hydrophilic
substrate. The strength and locality of the hydrogen bonding interactions
prohibit slip of the liquid over the substrate. The question then arises how
the contact line can advance during wetting. Using large-scale molecular
dynamics simulations we show that the contact line advances by single molecules
moving ahead of the contact line through two distinct processes: either moving
over or displacing other liquid molecules. In both processes friction occurs at
the molecular scale. We measure the energy dissipation at the contact line and
show that it is of the same magnitude as the dissipation in the bulk of a
droplet.
The friction increases significantly as the contact angle decreases, which
suggests suggests thermal activation plays a role. We provide a simple model
that is consistent with the observations.
\end{abstract}

\title{Molecular Origin of Contact Line Friction in Dynamic Wetting}
\author{Petter Johansson}
\author{Berk Hess}\email{hess@kth.se}
\affiliation{Department of Physics and Swedish e-Science Research Center,
    KTH Royal Institute of Technology,
    Stockholm, Sweden}
\date{\today}

\maketitle

\section{Introduction}
\noindent
The dynamics of wetting has been studied for more than a century.
The advent of new technological applications combined with advances in both
experimental and simulation techniques has led to increased attention in the
last decades \cite{DeGennes1985,Bonn2009,Snoeijer2013}.

\begin{figure}[b]
    \includegraphics{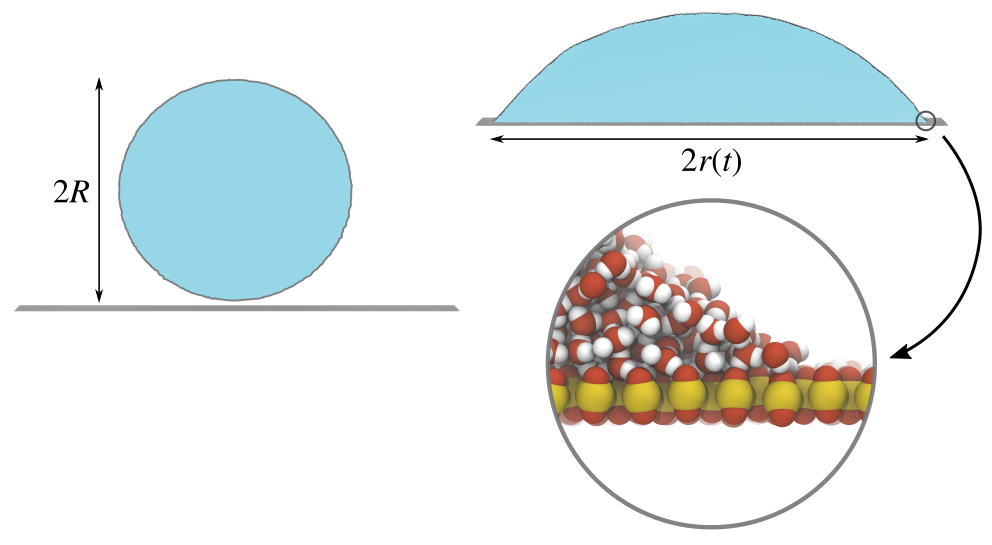}
    \caption{\label{fig:system-setup} The simulated systems are droplet
    cylinders with $R = 50$~nm, consisting of 1.2 million water molecules.
    After making contact with the hydrophilic substrates they spread out with
    a base radius $r(t)$. Zooming in shows the molecular nature of the
    droplets, particularly important at the contact line.
    }
\end{figure}

Droplet spreading on a substrate is commonly characterized by two stages:
an early stage where inertial forces
oppose droplet spreading \cite{Biance2004,Bird2008,Courbin2009} followed by a
later one where viscous forces dominate \cite{Heslot1989,Chen2014}.
Hydrodynamics along with a balance of surface tensions explains most details
of the wetting process. But there are still aspects which are not well
understood. In particular it is unclear what processes govern the
dynamics of the three-phase contact line between liquid, substrate and vapor
shown in figure \ref{fig:system-setup}. While hydrodynamical models can
incorporate these processes as boundary conditions they give no hint as to
what they are. A detailed, molecular picture is needed to understand what
is happening in the contact line region.

Two different situations can be distinguished in dynamic wetting.
The simpler case is a substrate--liquid
combination with nanoscale slip. Here there are no incompatibilities with
hydrodynamics \cite{Nakamura2013}. The more complicated case occurs with the
no-slip boundary condition, as is common when both liquid and substrate are
hydrophilic. In a continuum description, no slip is fundamentally incompatible
with contact line motion. Here we will focus on the no-slip case.

It has been shown that hydrodynamical models can reproduce experimental data by
accounting for the energy dissipation inside the bulk and at the contact line
\cite{Carlson2012}. The nature of this dissipation at the contact line is
unknown. Perrin \etal{}~relate it to the contact line pinning to nanoscopic
defects present at the substrate \cite{Perrin2016} but molecular simulation
studies have shown that it also appears on completely smooth, defect-free
substrates. In particular Nakamura \etal{}~show that the dissipation in their
system can be explained by liquid slip against the substrate
\cite{Nakamura2013}, but it has also been seen in similar smooth systems where
the liquid cannot slip over the substrate \cite{Johansson2015}. This
suggests that there is an additional contribution to contact line dissipation
due purely to how the contact line advances on a molecular level: a molecular
\emph{contact line friction} \cite{deRuijter99,Carlson2011}.
While it is likely that several or all of these modes of energy dissipation are
present in macroscopic wetting it is important to know where and how they
contribute.

A recent study has shown that the no-slip condition in dynamic wetting is
a consequence of liquid molecules forming \emph{hydrogen bonds} with
substrate molecules \cite{Johansson2015}. Hydrogen bonds are a particular
class of transient and non-covalent electrostatic bonds between polar
molecules, such as water.
They are characterized by being significantly stronger and more directional
than van der Waals interactions, but weaker than covalent bonds.
Because the strength of a hydrogen bond is an order of magnitude larger than
the thermal energy, these ``bonds'' localize the liquid molecules
to the underlying substrate molecules, thereby preventing slip.
Most atomistic studies of wetting have been performed with simple
Lennard-Jones type models for the substrate and often for the liquid as well
\cite{DeConinck2008,Ho2011,Winkels2012,Andrews2016}.
Simulations with models that capture hydrogen bonding have been limited
to smaller droplets \cite{Liu2010} or studied the effects of hydrogen
bonds within the liquid itself, not between liquid and substrate
\cite{Yuan2010,Ritos2013}. We will show that the dynamics of water spreading
on a hydrogen-bonding substrate is qualitatively different from Lennard-Jones
models, in particular at the contact line.

\begin{figure}
    \begin{subfigure}[b]{0.45\columnwidth}
        \centering
        \includegraphics{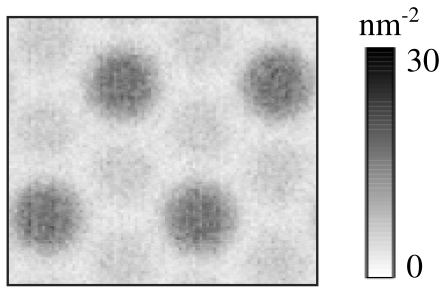}
        \caption{}
    \end{subfigure}
    \begin{subfigure}[b]{0.50\columnwidth}
        \centering
        \includegraphics[width=\columnwidth]{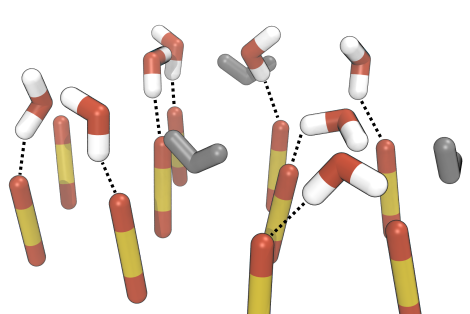}
        \caption{}
    \end{subfigure}
    \caption{\label{fig:hydrogen-bonds} a) A sampled 2D number density plot of
    hydrogen atoms in the first layer of water molecules shows that they are
    localized (hydrogen bonded) to the underlying \Sil{} ``molecules.'' b) A
    snapshot of hydrogen bonded molecules. All but the water molecules in gray
    are hydrogen bonded.
    }
\end{figure}

In this paper we use atomistic molecular dynamics (MD) simulations to measure
molecular contact line friction for a semi-realistic quasi-2D system of water
on silica with several hydrophilic static contact angles. The set-up is
detailed in figure~\ref{fig:system-setup}. Our system is large in order to
capture slow, long-term behavior and to provide sufficient statistics.
Using complex interactions between the
substrate and water we replicate replicate realistic
system set-ups and in particular the no-slip boundary
condition through the formation of liquid--substrate hydrogen bonds.
We observe that in no-slip wetting
the contact line friction is related to the dynamics of the
contact line. We propose a molecular explanation for this dissipation
and show that it matches our results.

\begin{figure*}[t]
    \centering
    \begin{subfigure}[b]{0.4\textwidth}
        \centering
        \includegraphics{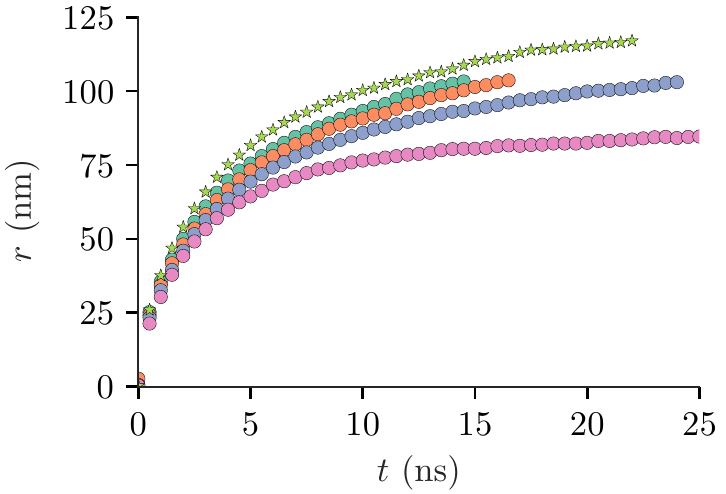}
    \end{subfigure}
    \;
    \begin{subfigure}[b]{0.55\textwidth}
        \centering
        \includegraphics{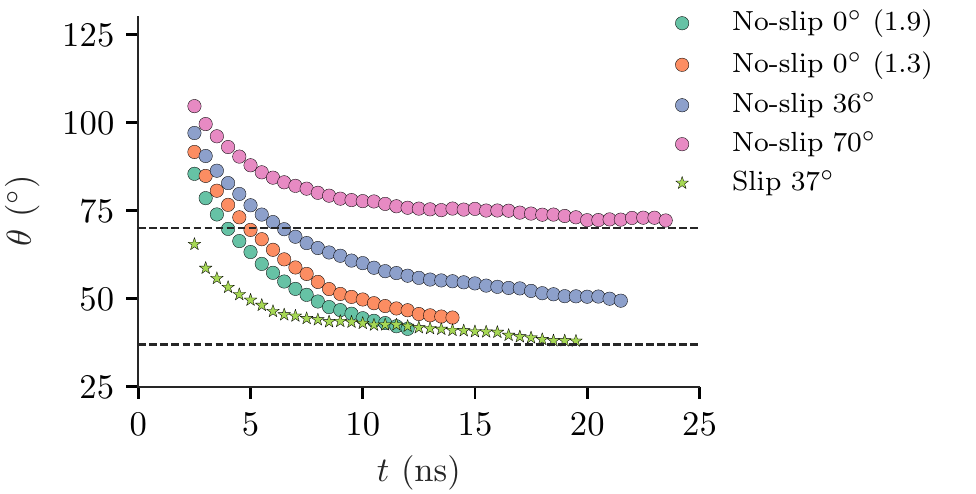}
    \end{subfigure}
    \caption{\label{fig:wetting-results} Measurements of the wetting radius
$r(t)$ (left) and the dynamic contact angle $\theta(t)$ (right) for simulations
on all involved substrates. The dashed lines denote the two equilibrium states
of angles $\mydeg{36}$ and $\mydeg{70}$.}
\end{figure*}

\section{System set-up}
\noindent
We use an idealized silica substrate to study hydrophilic wetting. It is
constructed as a monolayer of \Sil{} ``molecules'' consisting of two
negatively charged oxygen atoms and a positively charged silica atom
in the middle. These molecules do not have net charge nor a dipole moment,
so there is no long-range interaction with water. They are set in a triangular
formation with a spacing of 0.45~nm.

The polar nature of the substrate molecules give rise to strong, transient
hydrogen bonds with water molecules.
The strength of these short-ranged bonds result in a single, layer of
non-slipping water molecules ``bound'' to the substrate molecules.
Individual water molecules exchange between this layer and the bulk with
an average time of 24~ps for our systems. This creates
the no-slip condition between the substrate and the liquid, as the bottom
layer of molecules cannot move freely (figure \ref{fig:hydrogen-bonds}).

The wetting energetics of partially
wetting systems is related to the static contact angle $\theta_0$ through
Young's equation as $\gamma \cos{\theta_0} = \gamma_{SV} - \gamma_{SL}$ where
$\gamma \equiv \gamma_{LV}$, $\gamma_{SV}$ and $\gamma_{SL}$ respectively are
the liquid--vapor, solid--vapor and solid--liquid surface tensions. For
perfectly wetting systems $\gamma_{SV} - \gamma_{SL} \ge \gamma \cos \theta_0$.
By tuning the atomistic charges of our silica molecules we change the
substrate--liquid interactions to construct no-slip substrates with static
contact angles ranging from $\mydeg{0}$ to $\mydeg{70}$. To assert that the
water does not slip across the substrate, we use a Couette flow set-up. By
measuring the flow gradient we calculate that the slip length $\lambda$ for all
of our hydrophilic silica systems is $0.0 \pm 0.1$~nm.

We compare the results to a simpler system of water on a hydrophobic substrate
of Lennard-Jones atoms (see \cite{Johansson2015}). Since water cannot hydrogen
bond to this substrate, slip occurs. We measure a slip length of $1.1 \pm
0.1$~nm in an identical Couette flow set-up. Details of the characteristics of
all substrates are provided in table~\ref{tab:substrates}.

\begin{table}
    \centering
    \begin{ruledtabular}
        \begin{tabular}{lccccc}
            Substrate & $\qO{}$ ($e$) & $\lambda$ (nm) & $\theta_0$ ($^\circ$)
                & $\left ( \gamma_{SV} - \gamma_{SL} \right ) / \gamma$
                & $\fricCoeff{}$ ($\mu$) \\
            \hline
            Silica & $-0.84$ & $0.0 \pm 0.1$ & $0$ & $1.9$ & $110$  \\
            Silica & $-0.79$ & $0.0 \pm 0.1$ & $0$ & $1.3$ & $94$ \\
            Silica & $-0.74$ & $0.0 \pm 0.1$ & $36 \pm 0.2$
                & $\cos \theta_0 = 0.81$ & $76$ \\
            Silica & $-0.67$ & $0.0 \pm 0.1$ & $70 \pm 0.3$
                & $\cos \theta_0 = 0.34$ & $57$\\
            Atom. LJ & -- & $1.1 \pm 0.1$ & $37 \pm 0.4$
                & $\cos \theta_0 = 0.80$ & -- \\
        \end{tabular}
    \end{ruledtabular}
    \caption{\label{tab:substrates} Characteristics of the studied substrates.
The atoms of the silica molecules (\Sil{}) carry charges according to
$\qSi{} = -2 \qO{}$ where the oxygen charge $\qO{}$ is listed in the table, $e$ is the elementary charge. The atomistic LJ substrate atoms are non-charged
which results in water slipping across it. $\fricCoeff{}$ is a
substrate-dependent friction coefficient factor measured and described later
for the no-slip substrates.}
\end{table}

The water droplets are modeled as cylinders with an initial radius $R$ of 50 nm
and periodic width $w$ of 4.67 nm. They contain a total of 1.2 million water
molecules, or 3.6 million atoms, each. We chose SPC/E \cite{Berendsen87} as the
water model, which is characterized by its surface tension
$\gamma = \pow{5.78}{-2} \, \textrm{Pa} \, \textrm{m}$,
viscosity $\mu = \pow{8.77}{-4} \, \textrm{Pa} \, \textrm{s}$ and density
$\rho = 986 \, \textrm{kg} \, \textrm{m}^{-3}$. With the capillary speed
$v_\textrm{c} \equiv \gamma / \mu$ the Reynold's number of the system is
$\textrm{Re} \equiv \rho v_\textrm{c} R / \mu = 3.7$. See the appendix for
details on the simulation procedure.

\section{Simulation results}

\noindent
To study the dynamics of our wetting systems we sample our system in
two-dimensional bins during the simulation. Droplet interfaces are identified
using a simple density cutoff. Note that apart from the wetting radius $r$
(figure \ref{fig:system-setup}) all measurements presented here have been
center averaged over 5~ns of data to decrease the influence of thermal noise.

At each time $t$ the wetting radius $r(t)$ is taken as half of the droplet base
extension and the dynamic contact angle $\theta(t)$ is measured at a height of
1 nm above the substrate. These are shown in figure~\ref{fig:wetting-results}.
We see that the addition of slip allows for much faster relaxation towards the
equilibrium compared to the no-slip systems, even compared to the perfectly
wetting system. The no-slip systems have very similar dynamics in the early
inertial regime, diverging only as they transition into the viscous regime.
Over the simulations the no-slip substrate with a contact angle of $\mydeg{70}$
and the substrate with slip get close to their equilibrium states (dashed
lines) and their dynamics slow down dramatically.

Ideally one would like to measure this dissipation directly from the contact line-local shear stress.
However, this is in practice unfeasible using molecular simulations since
the thermal velocity of molecules is roughly 100 times higher than
the flow velocity (see Supplementary Material at [URL] for an extracted
shear field which has been averaged over 200 frames separated by 5~ps).
Thus we resort to modeling.
When considering the modes of energy dissipation during dynamic wetting the
bulk and interfacial dissipation terms are well characterized by continuum
models. The contact line dissipation rate is given by \cite{deRuijter99,Carlson2012a}
\begin{equation}
    \label{eq:contact-line-dissipation-rate}
    \clDiss \sim 2 w \mu_f v^2
\end{equation}
where $v = dr/dt$ is the contact line velocity, $2w$ is the total length of the
two contact lines and $\mu_f$ is referred to as a contact line friction
coefficient which has units of viscosity. Furthermore, in a damped regime
the contact line velocity is related to the force $\clForce{}$ acting
on the contact line through the friction coefficient (see \cite{Yue2011}
for a derivation and discussion in the context of the phase-field model):
\begin{equation}
    \label{eq:contact-line-friction-relation}
    v = \frac{\clForce{}}{\mu_f} =
        \frac{\gamma_{SV} - \gamma_{SL} - \gamma \cos{\theta}}{\mu_f} \, .
\end{equation}
This result is a direct relation between the contact line friction factor
$\mu_f$ and the two dynamic measurables $v(t)$ and $\theta(t)$, given the
system energetics. This allows us to measure the contribution of the contact
line friction to our dynamics.

\begin{figure}
    \centering
    \includegraphics{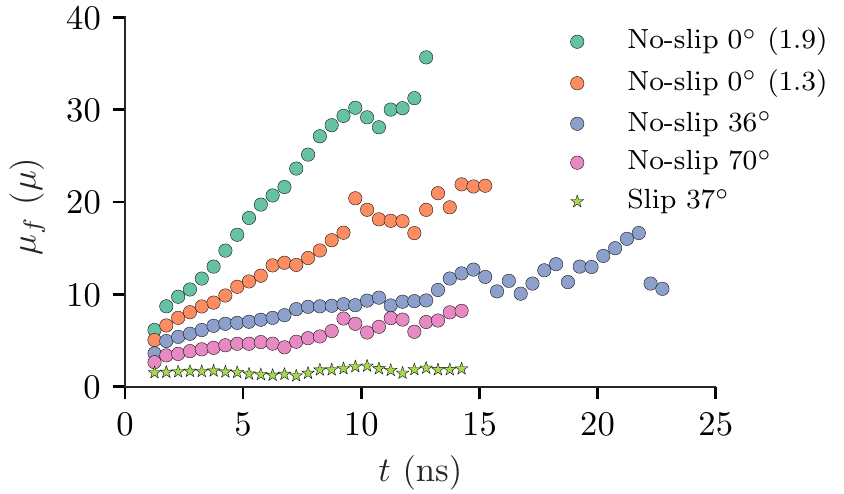}
    \caption{\label{fig:contact-line-dissipation} The measured contact line
friction coefficient $\mu_f$ in units of viscosity $\mu$ for all substrates.
When the liquid cannot slip the coefficient increases as the droplet wets the
substrate.}
\end{figure}

Figure \ref{fig:contact-line-dissipation} shows the contact line friction
coefficient $\mu_f$ calculated for each time and system using
\eqref{eq:contact-line-friction-relation}. There are a few things of note here:
The first and most obvious feature is that although $\mu_f$ has been thought of
as a constant which varies only between different substrates, we see that it
increases during the wetting process for our no-slip systems. Second is that
the system where slip is present has an apparent constant, non-zero friction
coefficient during its entire relaxation. Note that the data for the silica
with $\theta_0 = \mydeg{70}$ and the slip substrate is cut at 15~ns. This is
due to the systems being very close to equilibrium (figure
\ref{fig:wetting-results}), making the measurement of $v$ (and thus $\mu_f$)
dominated by thermal fluctuations.

\begin{figure}
    \centering
    \includegraphics{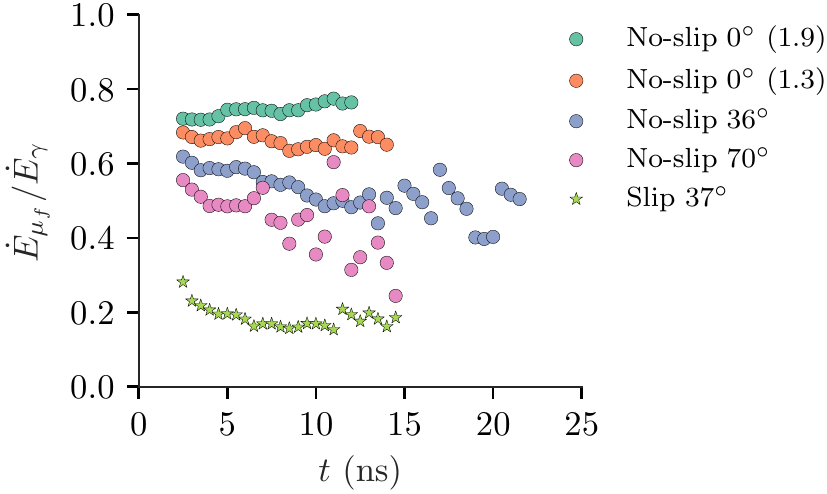}
    \caption{\label{fig:dissipation-versus-gain} Contact line dissipation rate
$\clDiss{}$ as a fraction of the energy gained as the droplet wets the
substrate $\eneGainRate{}$. For the no-slip substrates around half of the
gained energy is dissipated at the contact line.}
\end{figure}

To see whether or not the contact line friction is important to consider we use
\eqref{eq:contact-line-dissipation-rate} to calculate the dissipation rate and
compare it to the net energy $\eneGainRate{}$ being fed into the system as it
wets the substrate
\begin{equation}
    \eneGainRate{} = w \left [
            2 v \left ( \gamma_{SV} - \gamma_{SL} \right ) - \gamma \dot{s}
        \right ]
\end{equation}
where $\dot{s}(t)$ is the rate at which the length of the liquid--vapor
interface changes, which we measure from our simulations, and $w$ as before is
the interface (cylinder) width. A comparison between $\clDiss{}$ and
$\eneGainRate{}$ is shown in figure~\ref{fig:dissipation-versus-gain}. For the
no-slip substrates the dissipation at contact line accounts for more than half
of the energy gained as the droplet wets. Even for the substrate with slip, a
quarter of the gained energy is dissipated at the contact line. As such,
contact line friction is an important feature to consider when modeling similar
systems.
Note that the measurement error goes up heavily as $v$ and $\dot{s}$ approach
zero, which leads to a large variance for the no-slip system with $\mydeg{70}$
static contact angle, since it quickly approaches its equilibrium state.

\section{Energy dissipation in contact line friction}

\begin{figure*}
    \centering
    \includegraphics{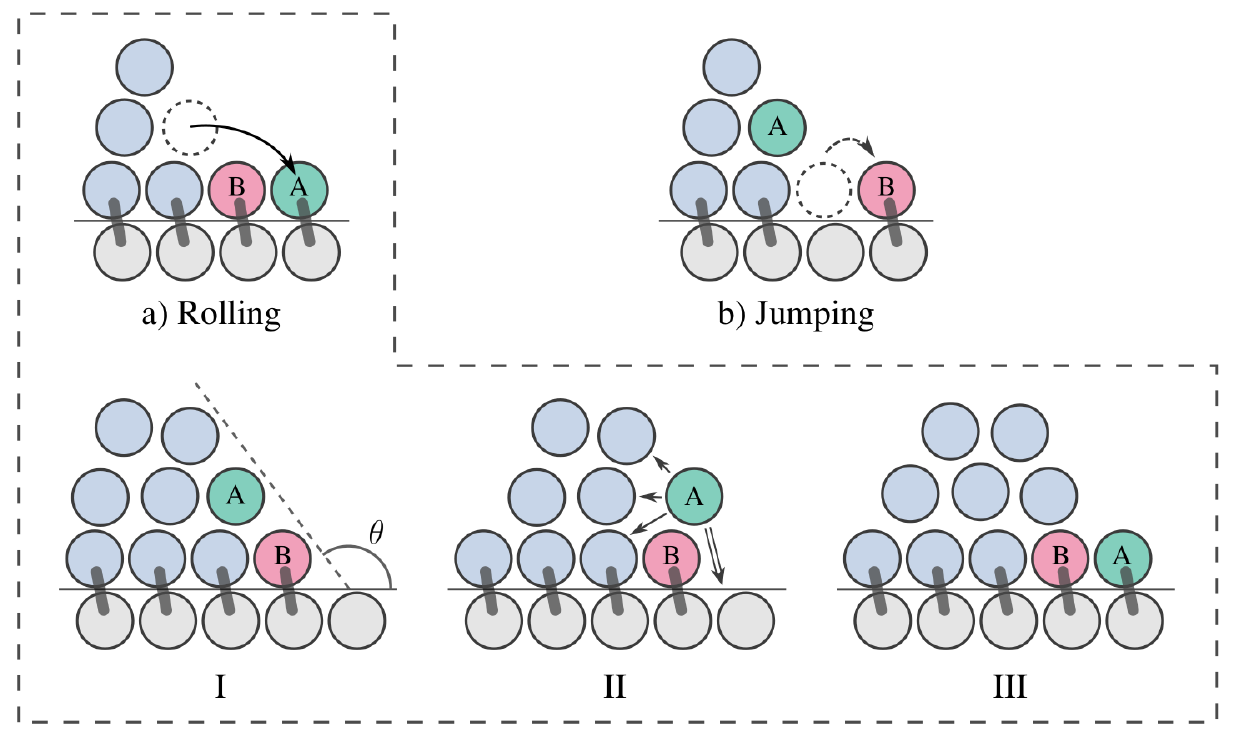}
    \caption{\label{fig:advancement-sketch} Modes of contact line advancement
for a no-slip system. The bottom layer of liquid molecules are hydrogen bonded to the
substrate. a) A molecule A rolls in from an upper layer to advance the contact
line. To advance from (I) it has to pass the intermediate state (II). Its
interactions with neighboring molecules lead to an energy barrier
that has to be crossed through a thermal fluctuation of the interface. After
crossing it, it is pulled towards the substrate (III).
    b) The outermost molecule B jumps to a neighboring lattice site. Described
by molecular kinetic theory.}
\end{figure*}

\begin{table}
    \centering
    \begin{ruledtabular}
        \begin{tabular}{ccccccc}
            $t$ (ns) & $r$ (nm) & $v$ (m/s) & $\theta$ & $\mu_f$ ($\mu$)
                & $f_\textrm{MKT}$ \\
            \hline
            $2.5$ & $50$ & $14$ & $\mydeg{95}$ & $4.4 \pm 0.5$
                & $0.19 \pm 0.01$ \\
            $8.0$ & $80$ & $3.0$ & $\mydeg{64}$ & $9 \pm 1$
                & $0.49 \pm 0.01$ \\
            $12.5$ & $90$ & $1.7$ & $\mydeg{55}$ & $12 \pm 2$
                &  $0.49 \pm 0.01$ \\
        \end{tabular}
    \end{ruledtabular}
    \caption{\label{tab:advancement-analysis} Data collected from simulations
on the no-slip $\mydeg{36}$ silica just in the switch from the inertial to
viscous wetting (at 2.5~ns) and at two stages of the viscous regime (at 8.0~ns
and 12.5~ns). The contact line speed $v$ decreases by an order of magnitude and
the fraction $f_\textrm{MKT}$ of molecules which advanced the contact line
through MKT-like jumps approaches 0.5 in the viscous regime.}
\end{table}

\noindent
Having shown that the contact line friction increases during the simulations of
our no-slip systems, we are naturally invited to consider why that is. This
question ties into the mechanism(s) of contact line advancement. For the case
of a no-slip system, we identify two important modes, sketched in figure
\ref{fig:advancement-sketch}. The first mode is a molecule from the non-bonded
layers of liquid molecules rolling onto the substrate from above (figure
\ref{fig:advancement-sketch}a). The second a molecule at the contact line
breaking its bond to the substrate and jumping to an adjacent potential well of
the substrate (figure \ref{fig:advancement-sketch}b). This is the base of
molecular kinetic theory (MKT) \cite{Blake1969}.

We measure which of these modes dominate during wetting using our molecular
trajectories. This is done by extracting the simulation state at points in the
simulation and replicating the system four times along the cylinder axis to
obtain more data and better statistics. We extract states from the no-slip
silica system with $\theta_0 = \mydeg{36}$ at three points: the first from when
the system transitions from the inertial regime to the viscous and the second
two at states within the viscous regime, when jumps between adjacent sites will
be more prominent. Data is collected over a period of 1~ns with a spacing of
1~ps. When a contact line advancement event is detected, the molecule which
advanced is back-traced in time. The fraction $f_\textrm{MKT}$ of which arrive
from an adjacent lattice site (corresponding to (b) in figure
\ref{fig:advancement-sketch}) is calculated from the results and reported in
table \ref{tab:advancement-analysis}. In the viscous regime the two described
modes contribute about equally to contact line advancement.

Let us consider the rolling mode (a) in more detail. Molecule A in figure
\ref{fig:advancement-sketch}a cannot drop down to the substrate without passing
molecule B, which acts as an obstacle. To pass this obstacle the molecule has
to move into a position from which it can drop down. The average thermal
velocity of a water molecule of 370~{m/s} is two orders of magnitude larger
than the local flow velocity (0.5--5~{m/s} in the viscous regime). Thermal
fluctuations dominate over an average mass flow in the vicinity of the contact
line. Thus molecule A will not simply roll into position to drop down: it has
to cross an energy barrier $\Delta E$. This barrier is due to the molecule not
wanting to break bonds with neighboring molecules. It has to cross this by
gaining sufficient momentum through a thermal fluctuation. After crossing the
barrier, the molecule will be pulled towards the substrate and form a hydrogen
bond with it.

At short range the interactions of atoms in the water molecule with the
substrate increase with Coulomb's law as $1/r$. Thus most of the gain in
kinetic energy occurs at the last moment, while the molecule is moving close to
vertically. As the molecule hydrogen bonds to the substrate, it suddenly stops.
This means that a large part of the vertical acceleration is not transferred to
subsequent water molecules. Rather those molecules bump into the molecules that
just advanced the contact line and mostly bounce back. Thus part of the gain in
surface energy is transformed into random thermal motion in the contact line
region. This heat will slowly dissipate into the environment. This is how
contact line friction arises when a hydrophilic liquid spreads on a smooth,
hydrophilic substrate.

\begin{figure}
    \centering
    \includegraphics{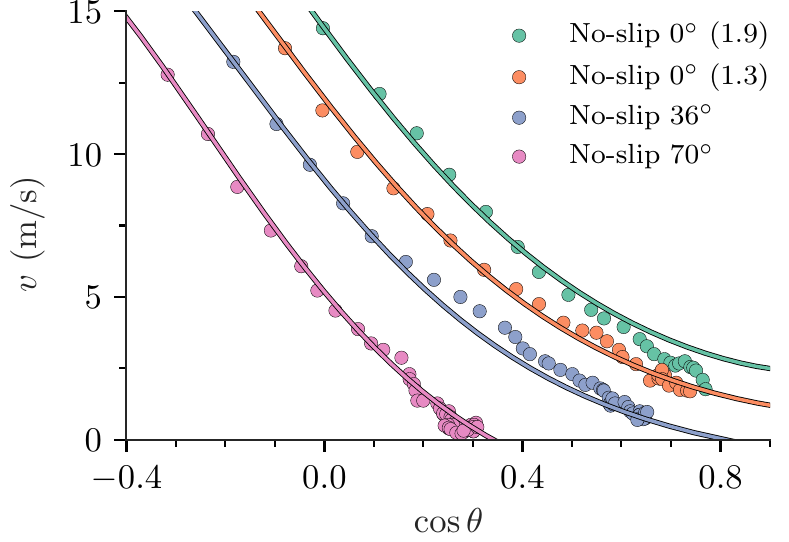}
    \caption{\label{fig:velocity-barrier-fit} The contact line friction modeled
as a geometry dependent energy barrier through the contact angle $\theta$. A
good agreement is reached between such a model (lines) and the measured
velocities (circles) in late-stage wetting for all bar the extremely
hydrophilic system. }
\end{figure}

We will now describe a simple model that explains why the contact line friction
coefficient increases with decreasing contact angle.
For the outermost molecule in the second layer to reach the transition state,
it needs to horizontally travel, on average, a distance of
$\Delta x = \tan{\thetadiff{}} + c$ water molecule diameters. In state (II) in
figure \ref{fig:advancement-sketch} the transition state is drawn as the water
molecule extending by half its diameter beyond the molecule at the contact
line. We take this state as the transition state and thus set $c = 0.5$ (note
that we cannot measure $c$ accurately).
The energy required to create such a fluctuation is given by the increase in
liquid--vapor interface area. This increase is proportional the square of the
displacement: $\eneBarrier{} \propto (\sin \theta \, \Delta x)^2$, where the
sine accounts for the difference in direction of $\Delta x$ and the normal of
the liquid--vapor interface. For a thermally activated process, the rate
decreases exponentially with the activation energy and we obtain (compare to
\eqref{eq:contact-line-friction-relation}):
\begin{equation}
    v = \frac{e^{-\eneBarrier{}}}{\fricCoeff{}} \, \clForce{}
        = \frac{e^{-a \left (
                    \sin{ \theta } [ \tan{ \thetadiff{} } + 0.5]
                \right )^2}}{\fricCoeff{}} \, \clForce{}
\end{equation}
where $\fricCoeff{}$ and $a$ are free model parameters.
We cannot fit all data using a single value for $\fricCoeff{}$. It is optimized
for each substrate. The best fit for this model is presented in figure
\ref{fig:velocity-barrier-fit} for our non-slip substrates with the values for
$\fricCoeff{}$ reported per-system in table \ref{tab:substrates} and the value
$a = 1.11$. We see that it provides a good prediction of the friction for most
systems. Only the most attractive substrate with
$(\gamma_{SV} - \gamma_{SL}) / \gamma = 1.9$ shows a deviation larger than the
statistical accuracy in the late stage.
The good fit suggests that the observed time dependence of the friction can be
explained by a dependence on the local geometry of the contact line.

Returning to the MKT-like mode, we can see how this fits into the model.
The MKT jump itself involves an energy barrier, but a forward
fluctuation of the second layer is also required to fill
the vacancy in the first layer. The amplitude of this fluctuation is about
one water molecule diameter less than the fluctuation for the rolling mode.
Since the activation energy of the MKT jump is approximately constant
and the activation energy of the second layer is proportional to the square
of the distance, the rolling mode will dominate at large contact angles,
while the MKT mode will increasingly dominate as the angle decreases.
This is exactly what we observe (see table \ref{tab:advancement-analysis}).

Finally we consider the substrate dependence of $\fricCoeff{}$.
In the early stages of wetting, when $\theta$ is larger than 90$^\circ$, the
energy barrier is negligible and the friction is dominated by $\fricCoeff{}$.
Here water molecules can be pulled directly to the substrate. As described
before, a significant part of the attraction occurs very close to the substrate
and most of that energy gain is lost in collisions.
The attraction between water and substrate, as given by
$\gamma_{SV} - \gamma_{SL}$, increases much faster than the charge difference.
Thus most of the difference in acceleration of water molecules to the different
substrates occurs at short distance with the substrate. Since a large part of
this final energy gain is dissipated, this explains why more attractive
substrate exhibit a larger $\fricCoeff{}$.

It is interesting to contrast these results to our slip substrate. In a slip
system the bottom layer is pulled along at similar speeds to the next layer.
The dissipation will to a large extent stem from the liquid slipping across the
substrate instead of from shear within the liquid wedge. The measurements shown
in figure~\ref{fig:contact-line-dissipation} are consistent with this picture.

\section{Conclusions}

\noindent
We have investigated contact line advancement of a hydrophilic liquid, water,
wetting a hydrophilic substrate. Due to the hydrogen bonding there is a no-slip
boundary condition at the substrate interface. In a Navier--Stokes setting this
is incompatible with a moving contact line. We have shown that the particle
nature of the liquid resolves this incompatibility at the length scale of a
single water molecule. Energy is dissipated in the processes that advance the
contact line and we have shown that this accounts for a substantial part of the
total energy dissipation in the viscous regime of wetting. The contact line
moves forward through a combination of two different modes: (a) from a molecule
rolling over the present contact line from an upper layer and (b) from one
contact line molecule jumping into an adjacent lattice site of the substrate.
This mode is MKT-like.
The rolling mode dominates in the initial phase of wetting, whereas both modes
occur with similar frequency in the viscous regime.

For the rolling mode, we have shown that molecules at the contact line form an
obstacle, because they are hard and of fixed, finite size. Passing such
obstacles requires a molecule, or group of molecules, to reach a sufficiently
high thermal activation energy. Most of this energy is lost through dissipation
after hydrogen bonding to the substrate. This results in additional energy
dissipation at contact line. We relate the energy barrier to the dynamical
geometry of the contact line and show that it yields good matches for the
velocity in late-stage wetting (figure \ref{fig:velocity-barrier-fit}).
Furthermore, we explain the difference in friction coefficient for differently
charged substrates. This is because the acceleration of water molecules to the
substrate is strongest at short distance, due to the $1/r$ dependence of the
electrostatic potential. After this final acceleration, water molecules are
locked in place when hydrogen bonding with the substrate. Other water molecules
cannot easily move around the locked molecules and a significant part of the
gain in convective kinetic energy is then dissipated.

All effects described above arise because water molecules are largely
immobilized when hydrogen bonding to a hydrophilic substrate. This is not the
case for Lennard-Jones type systems that are commonly used to model wetting. In
the latter case, liquid molecules can easily slip across the substrate which
produces an entirely different mode of contact line advancement and much lower
contact line friction. This is important to consider when modeling dynamic
wetting at the molecular scale.

\begin{acknowledgments}
We are grateful to Andreas Carlson for discussions leading to this publication.
This research was supported by the European Research Council (grant
no.~258980) and the Swedish Research Council (grant no.~2014-4505).
The simulations were performed on resources provided by the Swedish National
Infrastructure for Computing (SNIC) at the PDC Center for High Performance
Computing (SNIC 2015-10-30 and 2016-10-47). Analysis was performed and figures
created using \textsc{VMD} \cite{Humphrey1996} and \textsc{Matplotlib}
\cite{Hunter2007}.
\end{acknowledgments}

\appendix

\section*{\label{app:simulation-details} Appendix: Simulation set-up}

\noindent
Following our previous study \cite{Johansson2015}, the simulation systems
consist of a water droplet cylinder and an atomistic substrate. The no-slip
substrate consists of rigid \Sil{} tri-atomic ``molecules'' set in a hexagonal mono-layer.
For the substrate with slip, several layers of closely packed simple
Lennard-Jones atoms are used. These simple atoms are non-charged and do not
form hydrogen bonds with water molecules. See \cite{Johansson2015} for further
details. The initial radius $R$ of the water droplets are $50 \, \textrm{nm}$
and their width $w = 4.67 \, \textrm{nm}$. They consist of 1.2 million water
molecules. The droplet and the substrate are equilibrated to a temperature
$T = 300 \, \textrm{K}$ using a stochastic dynamics integrator. After this
stage a leap-frog MD integrator is used and wetting initiated by gently
bringing the droplet within the interaction range of the substrate. The water
droplet is not coupled to a thermostat during this stage of the simulation, but
a velocity rescaling thermostat is applied to the substrate to dissipate heat
with a time scale of 10~ps.

The water molecules are modeled using the SPC/E model. At 300~K this has a
viscosity of $8.77 \cdot 10^{-4}$ Pa\,s, a density of 986~{kg~m$^{-3}$} and a
surface tension of $5.78 \cdot 10^{-2}$~{Pa~m}. Note that the surface tension
is 20~\% slower than the experimental value. This is not an issue in this
study, as we study different contact angles and they depend on relative
differences between surfaces tensions.
For the \Sil{} atoms charges are varied to obtain different static contact
angles $\theta_0$. We measured the contact angles by letting a small droplet relax on
a substrate. The final equilibrium values are reported in table
\ref{tab:substrates}. The \Sil{} molecules are held in place by strong harmonic
potentials on both oxygen atoms. Short range interactions are treated fully up
to a cutoff of 0.9~nm and long range electrostatics using PME. Periodic
boundary conditions are applied along the axes transverse to the substrate,
along which the liquid spreads. Two repulsive potential walls are set along the
remaining sides to stop molecules from escaping the system. A time step of 2~fs
is used for the integrator.

Simulations were run using \Gromacs{} \cite{Abraham2015} in double precision on
the Beskow supercomputer at KTH. When running on 1280~cores a speed of
{5~ns/day} was achieved.

\end{document}